\documentclass[aps,prd,twocolumn,showpacs,preprintnumbers,amsmath,amssymb]{revtex4}

\usepackage{graphicx}
\usepackage{dcolumn}
\usepackage{bm,epsfig}

\begin{document}

\title{ $ND^{(*)}$ and $NB^{(*)}$ interactions in a chiral quark model}

\author{YANG Dan}
\author{LIU Jing}
\author{ZHANG Dan}
\affiliation{ School of Physical Science and Technology, Inner
Mongolia University, Hohhot 010021, China}

\begin{abstract}

$ND$ and $ND^*$ interactions become a hot topic after the
observation of new charmed hadrons $\Sigma_c(2800)$ and
$\Lambda_c(2940)^+$. In this letter, we have preliminary
investigated $S$-wave $ND$ and $ND^*$ interactions with possible
quantum numbers in the chiral SU(3) quark model and the extended
chiral SU(3) quark model by solving the resonating group method
equation. The numerical results show that the interactions between
$N$ and $D$ or $N$ and $D^*$ are both attractive, which are mainly
from $\sigma$ exchanges between light quarks. Further bound-state
studies indicate the attractions are strong enough to form $ND$ or
$ND^*$ molecules, except for $(ND)_{J=3/2}$ and $(ND^*)_{J=3/2}$ in
the chiral SU(3) quark model. In consequence $ND$ system with
$J=1/2$ and $ND^*$ system with $J=3/2$ in the extended SU(3) quark
model could correspond to the observed $\Sigma_c(2800)$ and
$\Lambda_c(2940)^+$, respectively. Naturally, the same method can be
applied to research $NB$ and $NB^*$ interactions, and similar
conclusions obtained, i.e. $NB$ and $NB^*$ attractive forces may
achieve bound states, except for $(NB^*)_{J=3/2}$ in the chiral
SU(3) quark model. Meanwhile, $S$ partial wave phase shifts of
$ND^{(*)}$ and $NB^{(*)}$ elastic scattering are illustrated, which
are qualitatively consistent with results from bound state problem..

\end{abstract}


\maketitle

Years of research works indicate that the chiral SU(3) quark model
\cite{xcqm-zhang} and the extended chiral SU(3) quark model
\cite{lrdai03} are successful in studying hadronic systems with
light flavors, such as baryon-baryon \cite{xcqm-zhang, lrdai03,
dai06, zd}, meson-baryon \cite{hfnk04, hfdk04, hfdk05, hfnk05},
meson-meson \cite{ophi, kpi}interactions, structures of multiquark
states \cite{si2,wu,wl07}, and so on. Recently, the chiral quark
model is applied to the heavy flavor sectors, and the tentative
works provide interesting results, which include masses of the
ground-state baryons \cite{zhqy}, the tetraquark states
\cite{hx07,hx08,zm08}, interactions of $DK$ \cite{ls}, $D\bar{D}$
and $B\bar B$ \cite{L}, $\Sigma_c\bar{D}$ and $\Lambda_c\bar{D}$
\cite{sgla}, and structures of $X(3872)$ \cite{x}, $Z_b(10610)$ and
$Z_b(10650)$ \cite{z} etc. Inspired by above successes, we are going
to investigate other interesting systems with heavy flavors by using
the chiral SU(3) quark model and the extended chiral SU(3) quark
model, which could provide another effective interpretation for new
hadrons with heavy flavors, and test the application of the chiral
quark model in the heavy flavor fields.

Belle and Babar Collaborations observed new states $\Sigma_c(2800)$
\cite{b} in 2005 and $\Lambda_c(2940)^+$ \cite{d} in 2007, and
subsequently they confirmed their achievements each other\cite{c,e}.
One of the probable explanations of their structures is
$\Sigma_c(2800)$ as a $ND$ and $\Lambda_c(2940)^+$ as a $ND^*$
molecular states\cite{h, i, f, g}. Then, it is worthwhile to study
$ND$ and $ND^*$ interactions dynamically with various methods to
further understand the nature of the $\Sigma_c(2800)$ and
$\Lambda_c(2940)^+$. In this Letter, we will primarily investigate
$S$-wave $ND$ and $ND^*$ interactions with possible quantum numbers
in the chiral SU(3) quark model and the extended chiral SU(3) quark
model by solving the resonating group method(RGM) equation. And the
same studies are performed to $NB$ and $NB^*$ systems to acquire
more information.

The chiral SU(3) quark model and the extended chiral SU(3) quark
model have been widely described in the literature \cite{xcqm-zhang,
lrdai03,dai06,zd,hfnk04,hfdk04,hfdk05,hfnk05}and we just give their
salient features here. The Hamiltonian of the
baryon($qqq$)-meson($Q\bar q$)($q$ means light quark and $Q$ heavy
quark) system can be written as
\begin{eqnarray}
H=\sum_{i}T_i-T_G+\sum_{i<j} V_{ij}\; ,
\end{eqnarray}
where $T_G$ is the kinetic energy operator for the c.m. motion, and
$V_{ij}$ represents the interactions of $qq$, $Q\bar q$, $Qq$ or
$q\bar q$. For $qq$ or $q\bar q$ pair,
\begin{eqnarray}
V_{qq(\bar q)}(ij)=V^{conf}(ij)+V^{OGE}(ij)+V^{ch}(ij).
\end{eqnarray}
For $Qq$ or $Q\bar q$ pair,
\begin{eqnarray}
V_{Qq(\bar q)}(ij)=V^{conf}(ij)+V^{OGE}(ij).
\end{eqnarray}
Note that as a preliminary study, the contributions of $q \bar q$
annihilation part and the Goldstone boson exchanges between the
heavy-light quark pairs are not
considered\cite{zhqy,hx07,hx08,zm08,ls,L,sgla}.

$V^{OGE}$ is the one-gluon-exchange (OGE) interaction, and the
confinement potential $V^{conf}$ is taken as linear form in this
work\cite{zhqy,hx07,hx08,zm08,ls,L,sgla}. $V^{ch}$ represents the
interaction from chiral field coupling, which includes scalar and
pseudoscalar boson exchanges in the chiral SU(3) quark model,
\begin{eqnarray}
V^{ch}(ij) = \sum_{a=0}^8 V_{\sigma_a}({\bm r}_{ij})+\sum_{a=0}^8
V_{\pi_a}({\bm r}_{ij}),
\end{eqnarray}
and also includes vector boson exchanges in the extended chiral
SU(3) quark model,
\begin{eqnarray}
V^{ch}(ij) = \sum_{a=0}^8 V_{\sigma_a}({\bm r}_{ij})+\sum_{a=0}^8
V_{\pi_a}({\bm r}_{ij})+\sum_{a=0}^8 V_{\rho_a}({\bm r}_{ij}).
\end{eqnarray}
Where $\sigma_{0},...,\sigma_{8}$ are the scalar nonet fields,
$\pi_{0},..,\pi_{8}$ the pseudoscalar nonet fields, and
$\rho_{0},..,\rho_{8}$ the vector nonet fields.

$V_{q(Q)\bar q}$ can be obtained from $V_{q(Q)q}$. For $V^{conf}$
and $V^{OGE}$, the transformation is given by $\lambda _i^c\cdot
\lambda _j^c\rightarrow -\lambda _i^c\cdot \lambda _j^{c*}$, while
\begin{equation}
V_{q\bar{q}}^{ch}=\sum_{j}(-1)^{G_j}V_{qq}^{ch,j}.
\end{equation}
Here $(-1)^{G_j}$ represents the G parity of the $j$th meson. The
detailed expressions can be found in Refs. \cite{xcqm-zhang,
lrdai03,dai06,zd,hfnk04,hfdk04,hfdk05,hfnk05}.

{\small
\begin{table}[htb]
\caption{\label{para} Model parameters for the light quarks. The
meson masses and the cutoff masses: $m_{\sigma'}=980$ MeV,
$m_{\epsilon}=980$ MeV, $m_{\pi}=138$ MeV, $m_{\eta}=549$ MeV,
$m_{\eta'}=957$ MeV, $m_{\rho}=770$ MeV, $m_{\omega}=782$ MeV,
$m_{\phi}=1020$ MeV, and $\Lambda=1100$ MeV for all mesons.}
\begin{center}
\begin{tabular}{cccc}
\hline\hline
  & $\chi$-SU(3)QM & \multicolumn{2}{c}{Ex. $\chi$-SU(3) QM}  \\
  &   I   &    II    &    III \\  \cline{3-4}
  &  & $f_{chv}/g_{chv}=0$ & $f_{chv}/g_{chv}=2/3$ \\
\hline
 $b_u$ (fm)  & 0.5 & 0.45 & 0.45 \\
 $m_u$ (MeV) & 313 & 313 & 313 \\
 $g_{ch}$    & 2.621 & 2.621 & 2.621  \\
 $g_{chv}$   & $-$   & 2.351 & 1.973  \\
 $m_\sigma$ (MeV) & 595 & 535 & 547 \\
\hline\hline
\end{tabular}
\end{center}
\end{table}}

The parameters for the light quarks are taken from the previous work
\cite{xcqm-zhang, lrdai03,dai06,zd}, which can give a satisfactory
description of the energies of the baryon ground states, the binding
energy of deuteron, the $NN$ scattering phase shifts, and $NY$ cross
sections. For simplicity, we only show them as Table \ref{para},
where the first set is for the chiral SU(3) quark model (I), the
second and third sets are for the extended chiral SU(3) quark model
by taking $f_{chv}/g_{chv}$ as 0 (II) and 2/3 (III), respectively.
Here $f_{chv}$ and $g_{chv}$ are the coupling constants for vector
coupling and tensor coupling of the vector meson fields,
respectively.

\begin{figure}[htb]
\epsfig{file=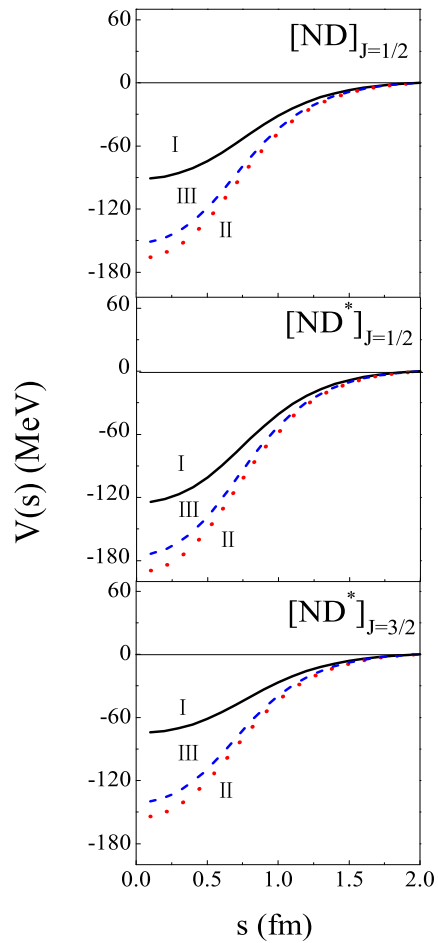,width=20.0cm,height=13.5cm} \vglue -0.8cm
\caption{\small \label{ham} The GCM matrix elements of the $S$-wave
$ND$ and $ND^*$  total effective potentials as functions of the
generator coordinate. The solid line represents the results obtained
in chiral SU(3) quark model with set I, and the dotted and dashed
lines represent the results in extended chiral SU(3) quark model
with set II and set III, respectively.}
\end{figure}

\begin{figure*}[htb]
\epsfig{file=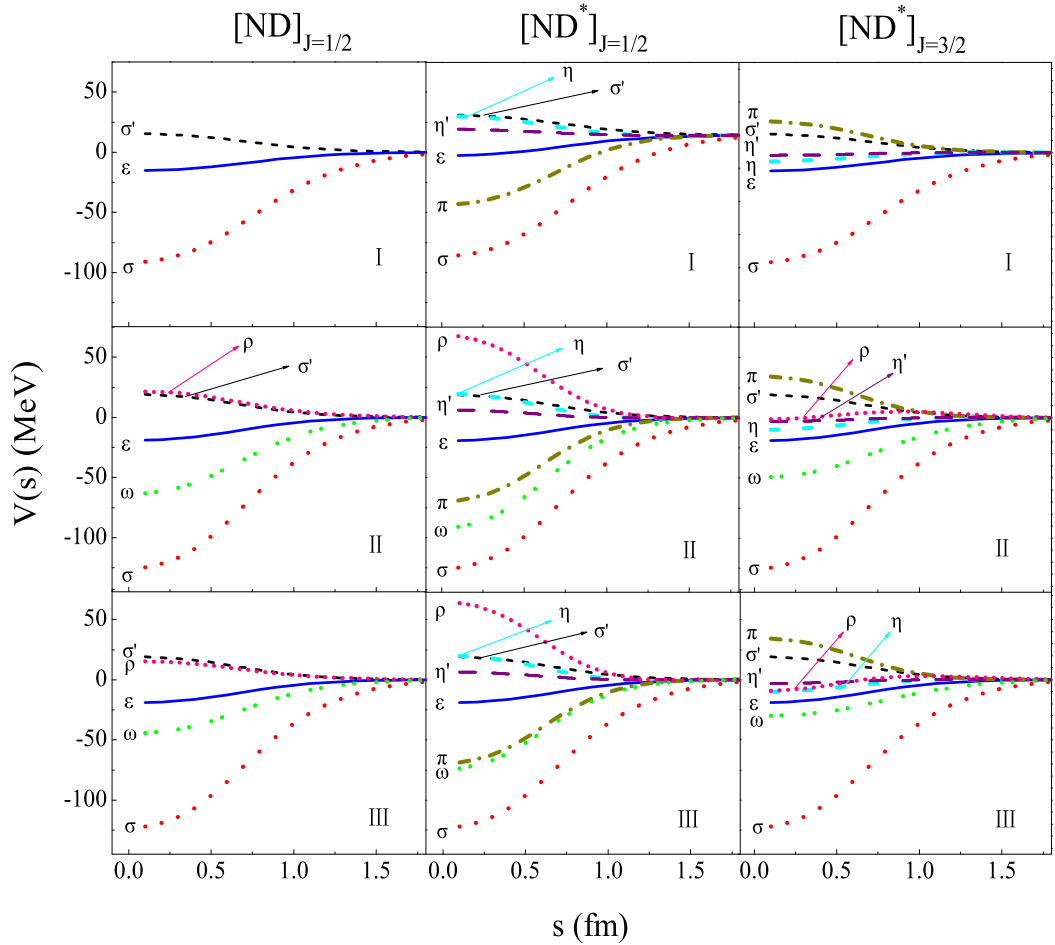,width=20.0cm,height=15.0cm} \vglue -1.6cm
\caption{\small \label{meson} The GCM matrix elements of the
$S$-wave $ND$ and $ND^*$ effective potentials from bosons exchanges
as a function of the generator coordinate.}
\end{figure*}

To examine the heavy quark mass dependence, we take several typical
values\cite{zhqy,sgla} $m_c=1430$ MeV\cite{hx07,hx08,zm08},
$m_c=1550$ MeV\cite{jv04}, $m_c=1870$ MeV\cite{bs93},$m_b=4720$
MeV\cite{hx07,hx08,zm08}, $m_b=5100$\cite{jv05}, $m_b=5259$ MeV
\cite{bs93}. While our calculated results indicate that the masses
of heavy quark ($m_Q$) contribute little effect, therefore we just
take $m_c=1430$ MeV and $m_b=4720$ MeV.

The OGE coupling constants and the confinement strengths for light
quarks can be determined by fitting the masses of ground-state
baryons with light flavors\cite{xcqm-zhang, lrdai03,dai06,zd},
whereas those for heavy quarks can be derived from the masses of
heavy mesons \cite{zhqy,hx07,hx08,zm08,ls,L,sgla,x,z}. Our
calculations suggest that between the two color-singlet clusters $N$
and $D^{(*)}$ or $B^{(*)}$, there is no OGE interaction and the
confinement potential. Therefore these values will not affect the
final results and we do not present them here.

With the parameters determined, the $S$-wave $ND^{(*)}$ and
$NB^{(*)}$ systems can be dynamically studied in the framework of
the RGM, a well established method for detecting the interaction
between two clusters. The details of solving the RGM equation can be
found in Refs. \cite{zd,hfnk04,hfdk04,hfdk05,hfnk05,ophi,kpi,o, p,
q, r}. By solving the RGM equation, one gets the energy of the
system, the relative motion wave function, and the elastic
scattering phase shifts.

\begin{table}[htb]
\caption{{\label{bind}}Binding energies of $ND^{(*)}$ and
$NB^{(*)}$.} \setlength{\tabcolsep}{2.6mm}
\begin {center}
\begin{tabular}{lccc}
\hline\hline   & $\chi$-SU(3)QM& \multicolumn{2}{c}{Ex. $\chi$-SU(3) QM}\\
\cline{2-4}
 & I & II & III \\
\hline
$(ND)_{J=1/2}$   &¡ª& 10.0&5.9\\
$(ND^*)_{J=1/2}$  &2.0& 18.1&12.8\\
$(ND^*)_{J=3/2}$  &¡ª& 6.6&3.2\\
\cline{1-4}
$(NB)_{J=1/2}$   &1.1& 18.2&12.8\\
$(NB^*)_{J=1/2}$  &7.7& 28.6&22.0\\
$(NB^*)_{J=3/2}$  & ¡ª& 13.7&8.9\\
 \hline\hline
\end{tabular}
\end{center}
\end{table}

\begin{figure}[htb]
\epsfig{file=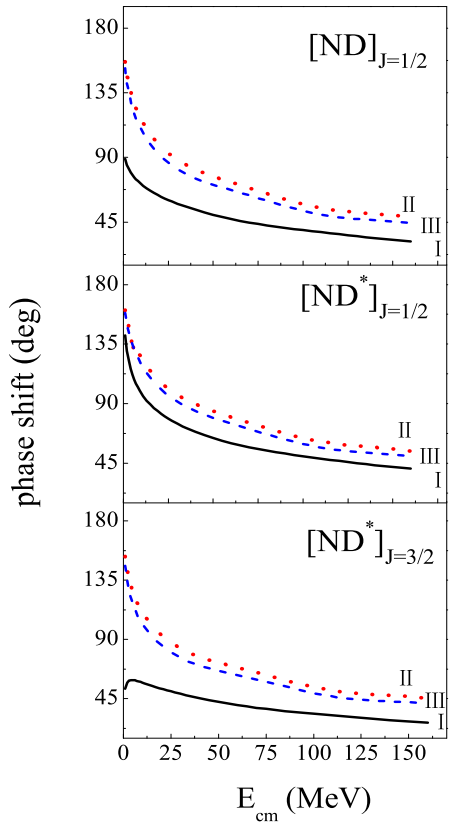,width=20.0cm,height=13.5cm} \vglue -1.7cm
\caption{\small \label{phase} $ND$ and $ND^*$ $S$-wave phase shifts
as functions of the energy of the center of mass motion. Same
notation as in Fig.\ref{ham}}
\end{figure}

Firstly, we consider $S$-wave $ND$ and $ND^*$ interactions.
Fig.\ref{ham} shows the diagonal matrix elements of the interacion
potentials for the $ND^{(*)}$ systems in the generator coordinate
method (GCM) \cite{o} calculations, which can be regarded as the
effective potentials of two clusters $N$ and $D^{(*)}$. In
Fig.\ref{ham}, $V(s)$ denotes the effective potentials between $N$
and $D^{(*)}$, and $s$ denotes the generator coordinate which can
qualitatively describe the distance between the two clusters. From
the figure, we can see interactions between $N$ and $D^{(*)}$ are
attractive, and the attractions in set II are greatest, while those
in set I are weakest. Further analysis shows that the attractions
dominantly come from $\sigma$ exchange between light quarks, and
others bosons exchanges nearly cancel each other (as shown in
Fig.\ref{meson}). And attractions from $\sigma$ exchanges in set II
are strongest.

In order to see whether the $ND^{(*)}$ attractions can result in
bound states or not, we have solved the RGM equation for a bound
state problem. The corresponding binding energies in three sets of
parameters are shown in Table \ref{bind}. One sees except that
$(ND)_{J=1/2}$ and $(ND^*)_{J=3/2}$ are unbound in the chiral SU(3)
quark model (set I), others can be weak bound states. Obviously,
$(ND)_{J=1/2}$ in set II with bing energy $10.0$ MeV and III with
$10.0$ MeV, and $(ND^*)_{J=3/2}$ in set II with $6.6$ MeV could
correspond to $\Sigma_c(2800)$ and $\Lambda_c(2940)^+$,
respectively.

The study of $ND^{(*)}$ elastic scattering processes has also been
performed by solving the RGM equation. The calculated S partial wave
 phase shifts of $ND^{(*)}$ are illustrated in Fig.\ref{phase} as a function
of $ND^{(*)}$ center of mass energy subtracted by the $ND^{(*)}$
threshold energy. We see that the signs of the phase shifts in these
two models(three sets of parameters) are the same, and the
magnitudes of the phase shifts in the extended chiral SU(3) quark
model are higher, especially for set II. This signify that the
$ND^{(*)}$ systems get more attractive interactions in the extended
chiral SU(3) quark model, which consists with the results of the
bind-state calculations (Fig.\ref{ham} and Table \ref{bind}).

The same research process is applied to $NB^{(*)}$ systems, and
similar conclusions are acquired. The binding energies are also
listed in Table \ref{bind}, where except that $(NB^*)_{J=3/2}$ is
unbound, others can be weak bound states. Effective interactions
between $NB^{(*)}$ are similar to Fig.\ref{ham} and \ref{meson}, and
S partial wave phase shifts are similar to Fig.\ref{phase} with all
curves move up about $0-10$ degree.

In summary, we have dynamically studied the interactions of $S$-wave
$ND^{(*)}$ and $NB^{(*)}$ system by solving RGM equation in the
chiral SU(3) quark model and the extended chiral SU(3) quark model,
including bound-state problem and elastic scattering phase shifts.
We have obtained some useful information. In our present
calculations the potentials between $N$ and $D^{(*)}$ or $B^{(*)}$
two clusters mainly come from $\sigma$ exchanges, which makes
$ND^{(*)}$ and $NB^{(*)}$ interactions are attractive. Furthermore,
such attractions are strong enough to form bound states except for
$(ND)_{J=1/2}$, $(ND^*)_{J=3/2}$ and $(NB^*)_{J=3/2}$ in the chiral
SU(3) quark model(set I). The information extracted from the $S$
partial phase shift of $ND^{(*)}$ and $NB^{(*)}$ is qualitatively
consistent with that from bound-state problem. In brief, the
observed $\Sigma_c(2800)$ and $\Lambda_c(2940)^+$ may be explained
as $ND$ molecular state with $J=1/2$ and $ND^*$ molecular state with
$J=3/2$, respectively .

\section*{Acknowledgments}

This project was supported by the Natural Science Foundation of
Inner Mongolia Autonomous Region of China (2015MS0115).


\begin{thebibliography}{99}
\bibitem{xcqm-zhang}Zhang Z Y, Yu Y W, Shen P N, Dai L R, Faessler A and Straub U 1997 Nucl. Phys. A 625 59
\bibitem{lrdai03}Dai L R, Zhang Z Y, Yu Y W, and Wang P 2003 Nucl. Phys. A 727, 321
\bibitem{dai06}Dai L R, Zhang Z Y and Yu Y W 2006 Chin. Phys. Lett. 23 3215
\bibitem{zd}Zhang D, Huang F, Dai L R,  Yu Y W, and Zhang Z Y 2007 Phys. Rev. C 75 024001
\bibitem{hfnk04}Huang F, Zhang Z Y and Yu Y W 2004 Phys. Rev. C 70 044004
\bibitem{hfdk04}Huang F and Zhang Z Y 2004 Phys. Rev. C 70 064004
\bibitem{hfdk05}Huang F and Zhang Z Y 2005 Phys.Rev. C 72 068201
\bibitem{hfnk05}Huang F and Zhang Z Y 2005 Phys. Rev. C 72 024003
\bibitem{ophi}Wang W L, Huang F, Zhang Z Y, and  Liu F 2010 Mod. Phys. Lett. A 25 1325
\bibitem{kpi}Huang F, Zhang Z Y, and Yu Y W 2005 Commun. Theor. Phys. 44 665
\bibitem{si2}Huang F, Zhang Z Y, Yu Y W, and Zou B S 2004 Phys. Lett. B 586 69
\bibitem{wu}Zhang D, Huang F, Zhang Z Y, and Yu Y W 2005 Nucl. Phys. A 756 215
\bibitem{wl07}Wang W L, Huang F, Zhang Z Y, Yu Y W and Liu F 2007 J. Phys. G 34 1771
\bibitem{zhqy}Zhao Q Y, Zhang D, and Zhang Q Y 2011 Chin. Phys. Lett. 28 071201
\bibitem{hx07}Zhang H X, Zhang M, and Zhang Z Y 2007 Chin. Phys. Lett. 24 2533
\bibitem{hx08}Zhang H X, Wang W L, Dai Y B and Zhang Z Y 2008 Commun. Theor. Phys. 49 414
\bibitem{zm08}Zhang M, Zhang H X, and Zhang Z Y 2008 Commun. Theor. Phys. 50 437
\bibitem{ls}Zhang D, Zhao Q Y, and Zhang Q Y 2009 Chin. Phys .Lett. 26 091201
\bibitem{L}Li M T, Wang W L,Dong Y B, and Huang F,  2012 Int. J. Mod. Phys. A 27 1250161
\bibitem{sgla}Wang W L, Huang F, Huang F, and Zou B S 2011 Phys. Rev. C 84 015203
\bibitem{x}Liu Y R and Zhang Z Y 2009 Phys. Rev. C 79 035206
\bibitem{z}Li M T, Wang W L,Dong Y B, and Zhang Z Y 2013 J. Phys. G 40 015003
\bibitem{b}R. Mizuk et al. [Belle Collaboration] 2005 Phys. Rev. Lett. 94 122002
\bibitem{d}B. Aubert et al. [Babar Collaboration] 2007 Phys. Rev. Lett. 98 012001
\bibitem{c}B. Aubert et al. [Babar Collaboration] 2008 Phys. Rev. D 78 112003
\bibitem{e}R. Mizuk et al. [Belle Collaboration] 2007 Phys. Rev. Lett. 98 262001
\bibitem{h}He J, Ye Y T, Sun Z F, and Liu X 2010 Phys. Rev. D 82 114029
\bibitem{i}P. G. Ortega, D. R. Entem, and F. Fernandez 2013 Phys. Lett. B 718 1381
\bibitem{f}Zhang J R 2014 Phys. Rev. D 89 096006
\bibitem{g}Zhang J R 2014 Int. J. Mod. Phys. Conf. Ser. 29 1460220
\bibitem{jv04}Vijande J, Garcilazo H, Valcarce A and Fernandez F 2004 Phys. Rev. D 70 054022
\bibitem{bs93}Silvestre-Brac B and Semay C 1993 Z Phys. C 57 273
\bibitem{jv05}Vijande J, Fernandez F and Valcarce A 2005 J. Phys. G 31 481
\bibitem{o}Wildermuth K and Tang Y C 1977 A Unified Theory of the Nucleus (Vieweg, Braunschweig)
\bibitem{p}Kamimura M 1977 Suppl. Prog. Theor. Phys. 62 236
\bibitem{q}Oka M and Yazaki K 1981 Prog. Theor. Phys. 66 556
\bibitem{r}Straub U, Zhang Z Y, Brauer K, Faessler A, Kardkikar S B and Lubeck G 1988 Nucl. Phys. A 483 686
\end{thebibliography}
\end{document}